\documentclass[11pt]{article}
\pdfoutput=1
\usepackage{amsmath}
\usepackage{amssymb}
\usepackage{graphicx,bbm,mathrsfs}
\usepackage{nicefrac}
\usepackage{bbm}
\usepackage{geometry}       
\usepackage{jheppub}       

\def\ie{{\it i.e.}}

\def\fbi{~{\rm fb}^{-1}}

\newcommand{\be}{\begin{equation}}  
\newcommand{\ee}{\end{equation}}  
\newcommand{\bea}{\begin{eqnarray}}  
\newcommand{\eea}{\end{eqnarray}}

\renewcommand{\O}{\mathcal O}

\newcommand{\WB}{W\hspace{-1.5pt}B}
\newcommand{\WW}{W\hspace{-.5pt}W}
\newcommand{\BB}{B\hspace{-.5pt}B}

\newcommand\lsim{\mathrel{\rlap{\lower4pt\hbox{\hskip1pt$\sim$}}
    \raise1pt\hbox{$<$}}}
\newcommand\gsim{\mathrel{\rlap{\lower4pt\hbox{\hskip1pt$\sim$}}
    \raise1pt\hbox{$>$}}}

\newcommand{\captionfonts}{\small}

\newcommand{\approptoinn}[2]{\mathrel{\vcenter{
  \offinterlineskip\halign{\hfil$##$\cr
    #1\propto\cr\noalign{\kern2pt}#1\sim\cr\noalign{\kern-2pt}}}}}

\makeatletter  
\long\def\@makecaption#1#2{%
  \vskip\abovecaptionskip
  \sbox\@tempboxa{{\captionfonts #1: #2}}%
  \ifdim \wd\@tempboxa >\hsize
    {\captionfonts #1: #2\par}
  \else
    \hbox to\hsize{\hfil\box\@tempboxa\hfil}%
  \fi
  \vskip\belowcaptionskip}
\makeatother   

\begin{document}

\vspace*{1.2cm}

\begin{center}

\thispagestyle{empty}

{\Large\bf Probing the scale of New Physics at the LHC: the example of Higgs data }\\[10mm]

\renewcommand{\thefootnote}{\fnsymbol{footnote}}

{\large Sylvain~Fichet$^{\,a}$\footnote{sylvain.fichet@lpsc.in2p3.fr}}\\[10mm]

\addtocounter{footnote}{-1} 

{\it
$^{a}$~International Institute of Physics, UFRN, 
Av. Odilon Gomes de Lima, 1722 - Capim~Macio - 59078-400 - Natal-RN, Brazil \\
}

\vspace*{12mm}

{  \bf  Abstract }
\end{center}

\noindent 
We present a technique to  determine the scale of New Physics (NP) compatible with any set of data, 
relying on  well-defined  credibility intervals.   
Our approach relies on the statistical view of the effective field theory capturing New Physics at low energy.
We introduce formally the notion of testable NP and show that it ensures integrability of the posterior distribution.
We apply our method to the Standard Model Higgs sector in light of recent LHC data, considering two generic scenarios.  
In the scenario of democratic higher-dimensional operators generated at one-loop, we find the testable NP scale to lie within $[10,260]$ TeV at $95\%$ Bayesian credibility level.
 In the scenario of loop-suppressed field strength-Higgs operators, the testable NP scale is  within $[28,1200]$ TeV at $95\%$ Bayesian credibility level.
  More specific UV models are necessary to allow lower values of the NP scale.

\vspace*{1cm}

\clearpage
%

\section{Introduction}  \label{se:introduction}

Several major experimental and theoretical facts like the measurement of neutrino masses, proofs of the existence of dark matter, as well as the hierarchy problem or the striking hints for Grand Unification all point towards the existence  of physics beyond the Standard Model (SM).  Although there are strong expectations that such New Physics (NP) should show up at an energy scale close to the electroweak scale, direct searches for new states have so far turned out to be unsuccessful. Indirect constraints from electroweak precision measurements at LEP also push the NP scale $\Lambda$ above the electroweak scale. 

Overall, it seems that $\Lambda$ should be substantially higher than the electroweak scale, $\Lambda \gg m_Z$. 
This paradigm is adopted in a large amount of propositions of new physics. We adopt this fairly general hypothesis  in the present work.
 It implies that the NP involved in physical processes at an energy scale $E \ll \Lambda$ can be integrated out. This results in a low-energy effective theory, consisting of the Standard Model supplemented by infinite series of local, higher-dimensional operators (HDOs) involving negative powers of the NP scale $\Lambda$, 
\be
\mathcal{L}_{\rm eff}=\mathcal{L}_{\rm SM}+\sum_{i,m} \frac{\alpha_{i}}{\Lambda^{m_i}} \mathcal{O}_{i}\,. \label{L_eff}
\ee

Considering a set of experimental observations through this effective description of new physics, 
we can wonder what information can be obtained about $\Lambda$.  For a dataset perfectly compatible with the SM, it is common to derive a lower bound on $\Lambda$, barring some fine-tuned cancellations among HDO-induced contributions. On the other hand, if data show a deviation with respect to the SM, arbitrary high values of $\Lambda$ should be also disfavoured, as the effective theory reduces to the SM  in the decoupling limit  $\Lambda\rightarrow\infty$ and cannot explain the discrepancy.
Finding a general method to consistently infer the range of $\Lambda$ compatible with some data -- whether they deviate or not from the SM -- is the subject of the present work.

We are going to use the effective theory approach within the framework of Bayesian statistics. 
An important feature of the Bayesian framework  is that any irrelevant parameter can be consistently eliminated in a well-defined way through integration. Here we will be mainly interested in  the probability distribution of $\Lambda$, $p(\Lambda | \rm data)$, which will be obtained through integration over all the $\alpha_i$ coefficients. Adopting a Bayesian view  is appropriate to account for the  generic  character of the scenario we will consider (\ie~it ensures that no fine-tuning is present in the scenario). 
\footnote{ Adopting such viewpoint already provided useful tools to treat anarchical models of the SM flavour sector, see \cite{Brummer:2011cp}.}


The outline of this note is as follows. In Section \ref{se:EFT_inference} we shortly review the basics of Bayesian inference and discuss its application to effective theories.  In Section \ref{se:consistent_framework} we show that one has to require NP to be testable to obtain an integrable posterior. The basic MCMC setup and conceptual subtleties inherent to our approach are discussed in Section \ref{se:MCMC}. 
Although inference on $\Lambda$ applies to any kind of data, it is particularly motivated by current LHC results.  In Section \ref{se:Higgs} we  apply our method to the Standard Model Higgs sector, using the latest pieces of information available from CMS, ATLAS and Tevatron. We discuss the leading constraints and the necessary conditions favouring lower values of the NP scale.


\section{Effective theory and Bayesian inference} \label{se:EFT_inference}
Let us briefly review necessary notions of Bayesian statistics (see \cite{Trotta:2008qt} for an introduction).
 In this approach, the notion of probability $p$ is defined as the degree of belief about a proposition.
Our study lies in the domain of Bayesian inference, which is based on the relation
\be
p(\theta |d,\mathcal{M})\propto p(d|\theta,\mathcal{M})p(\theta|\mathcal{M})\,.
\ee
In our case $\theta\equiv\{\Lambda, \alpha_{1\ldots n}\}$ are the parameters of the higher-dimensional operators (HDOs) defined in Eq. \eqref{L_eff}.
The parameter space will be denoted by $\mathcal{D}$.
 \footnote{Notice in general $\mathcal{D}$ should also enclose the SM parameters. However this is not relevant for the present work. In the Higgs sector study we will perform, modifications of the SM parameters do appear, but they can  always be expressed in terms of the HDO parameters.  }
  $\mathcal{M}$ is the Standard Model extended with HDOs, and $d$ represents the experimental data.
  The distribution $p(\theta |d,\mathcal{M})$ is the so-called posterior distribution, $p(d|\theta,\mathcal{M})\equiv L(\theta)$ is the likelihood function encoding experimental data, and  $p(\theta|\mathcal{M})$ is the prior distribution, which represents our a priori degree of belief in the parameters.

The posterior distribution is the core of our results. Being interested in the new physics scale, we focus on the marginal posterior $p(\Lambda|d,\mathcal{M})$, obtained by integrating all HDO coefficients $\alpha$'s, 
\be p(\Lambda|d,\mathcal{M})\propto\int d^n\alpha_i\, p(\Lambda,\alpha_i|\mathcal{M})L(\Lambda,\alpha_i)\,.\ee

The prior and posterior distributions do not need to be normalized to unity to carry out the  inference process in its broader meaning. For example, assuming some significant deviation from the SM is present in the data, it is sufficient to look at the bump in $\Lambda$'s improper posterior to have a good idea about the values of $\Lambda$ favored by the data. 
However, to go further and determine intervals associated with an actual probability (Bayesian Credible intervals), the posterior does need to be normalizable.   More precisely, the posterior needs to be ``proper''. It should be integrable on a unbounded domain like $\mathbb{R}$. Over a bounded domain, the integral should  be independent of the bound of the domain, unless the bound is well justified.

 In the rest of this section we will observe that the $\Lambda$'s posterior is improper. We will find the conceptual subtlety at the origin of this improperness, then propose a slight conceptual change leading to a proper $\Lambda$'s posterior.
In this work we  consider as valuable the ability to determine  Bayesian Credible intervals, and thus to have proper posteriors. However, even without paying particular interest to properness and Bayesian Credible intervals, 
the conceptual observations and their consequences that we will present below are in any case relevant for anyone interested in inference on $\Lambda$.




For concreteness, we give to the NP scale a logarithmically uniform distribution,
\be p(\Lambda)\propto \frac{1}{\Lambda}\,. \label{prior_lambda}\ee
By doing so, all the orders of magnitude are given the same probability weight.
This is arguably the most objective choice, justified by the ``principle of indifference'' \cite{press,jaynes}. \footnote{The ``principle of indifference'' maximizes the objectiveness of the priors. Once a transformation law $\gamma =f(\theta)$ irrelevant  for a given problem is identified, this principle provides the most objective prior by identifying $p_\Theta\equiv p_\Gamma$ in the relation $p_\Theta(\theta)d\theta=p_\Gamma(\gamma)d\gamma$. } 
There is no sensible argument to fix the upper bound on $\Lambda$. The prior of $\Lambda$ is therefore improper.

Similarly, we give uniform priors to the $\alpha$'s. 
Contrary to the domain of $\Lambda$, there are well justified bounds on $\alpha$'s because of perturbativity of the EFT approach. Indeed for $\Lambda>4\pi v$, perturbativity implies $\alpha_i\in[-16\pi^2,16\pi^2]$. 
We refer to \cite{Dumont:2013wma} for more details about the bounds on HDOs coefficients.
Although the priors adopted above are well motivated, the whole approach including the upcoming statements remains valid for any kind of priors,  as long as the domain of $\alpha$'s remains bounded.


Determining the posterior distribution of $\Lambda$ is a standard Bayesian procedure. However a peculiarity of the $\Lambda$ posterior is that in the decoupling limit $\Lambda\rightarrow \infty$, the likelihood tends to its SM value $L\rightarrow L^{SM}$. As the logarithmic prior of $\Lambda$ is also improper, it turns out that the posterior distribution is improper in the $\Lambda$ direction,
\be
p(\Lambda\rightarrow\infty|d)\propto \frac{L_{SM}}{\Lambda}\,.
\ee

To understand the origin of this improperness, let us introduce the notion of ``testability", carrying the  usual meaning as given e.g. in philosophy of science (see e.g. \cite{kuhn1996structure}). Considering the effective Lagrangian Eq. \eqref{L_eff}, we observe that, for $\Lambda\rightarrow\infty$, the new physics cannot manifest itself in the data. It is therefore not testable at $\Lambda\rightarrow\infty$.
 However, the behaviour of $p(\Lambda|d)$ in the decoupling limit does not seem to reflect this fact, as it remains constant up to the $1/\Lambda$ factor coming only from the prior.

Let us  be more precise by translating the notion of testability in a formal way. We adopt the following definition as a Bayesian translation of  testability.
 ``A model is testable with respect to the SM for a given dataset $d$ whenever $L\neq L^{SM}$".
We would like to know what happens to our posterior when we require testability.
For a continuous parameter space, requiring testability corresponds to excising a slice $\Omega_{\Lambda,L^{SM}}$ of the parameter space, defined as  $\Omega_{\Lambda,L^{SM}}=\{\alpha_i| L(\Lambda,\alpha_i)= L^{SM} \}$. 
Therefore, by requiring testability of the HDOs-extended SM, inference is made on the possibilities of new physics which are actually testable by the data.
Stated differently, to the initial question ``What can we learn about  $\Lambda$ from $d$ ?'', we already know that the answer is ``Nothing'' whenever $L=L^{SM}$. We therefore discard this particular possibility, to investigate the NP which can be actually probed by $d$.

The fact that the requirement of testability leads to a proper posterior will be demonstrated in Sect. \ref{se:consistent_framework} and in the Appendix. We admit it for the rest of this section.
Requiring testability, the marginal posterior of the NP scale $\Lambda$ is then expressed as
\be p^*(\Lambda|d,\mathcal{M})=p(\Lambda|L\neq L^{SM},d,\mathcal{M})\,.\label{p_Lambda_tilde}
\ee
In our approach this distribution  is the relevant object to inform us on the NP scale and will be therefore at the center of our interest for subsequent applications.  We refer the reader to Sect. \ref{se:consistent_framework}  for a formal discussion.

Notice this subtlety about testability usually does not matter in cases where the posterior is proper. Typically, the likelihood is continuous and bounded, such that the subdomain $\Omega_{\Lambda,L^{SM}}$ has  measure zero. Excluding this subdomain therefore does not change integrals of the posterior, and leaves the results of inference unchanged. The requirement of testability becomes important in our case because the posterior is improper. More generally this problem is susceptible to appear whenever the NP scale is a free parameter of the model.

Some qualitative comments can be made about the different effects driving the shape of the $ p^*(\Lambda|d,\mathcal{M})$ posterior. Both tails will drop to zero,  fast enough to let the distribution be integrable.
Let us consider the low--$\Lambda$ tail of the distribution.
Even though experimental constraints push $\Lambda$ to high values, it often happens that some precise cancellations between various HDOs contributions allow $\Lambda$ to go to low values. However, the regions of parameter space in which precise cancellations occur have weak statistical weight by construction, such that their unnatural character is built-in the Bayesian approach (see \cite{Fichet:2012sn} for more considerations on naturalness). We conclude that the low--$\Lambda$ tail is set by the trade-off between goodness-of-fit and possible fine-tuning. 
Considering the high-$\Lambda$ tail, if the data $d$ are compatible with the SM, the shape is asymptotically independent of $d$, and is only dictated by the probability of $\mathcal{M}$ to be testable. In contrast, when $d$ shows some deviation with respect to the SM, a good fit of the deviation  favours low values of $\Lambda$. The high--$\Lambda$ tail is thus shaped by the two effects. It is set by default to a profile depending only on $\mathcal{M}$, which is overwhelmed by the shape dictated by goodness-of-fit once an excess appears in the data. The high-$\Lambda$ tail behaviour can be observed in the toy model of App. \ref{app:BSM_coin}.

\section{Inference on testable new physics} \label{se:consistent_framework}

In this Section we scrutinize the posterior to better understand how its integral diverges. We then show that the requirement of having testable NP  leads to a proper posterior. 
It is necessary to use the framework of Lebesgue integration to treat rigorously the following questions. In doing so, we will introduce the Lebesgue measure $\mu$. \footnote{This is the appropriate measure for continuous probability spaces. For brevity we will omit the argument of the integrand when no ambiguity is possible. We are going to use the extended real set $\bar{\mathbb{R}}=\mathbb{R}\cup \{-\infty,\infty\}$.}
In what follows we let $\Lambda$ go to infinity, such that the designation``proper'' is equivalent to ``integrable''.  Various proofs are reported in the Appendix, as well as a useful example of explicit computation within a toy model.

\begin{figure}
	\centering
		\includegraphics[trim=0cm 0cm 0cm 0cm, clip=true,width=6cm]{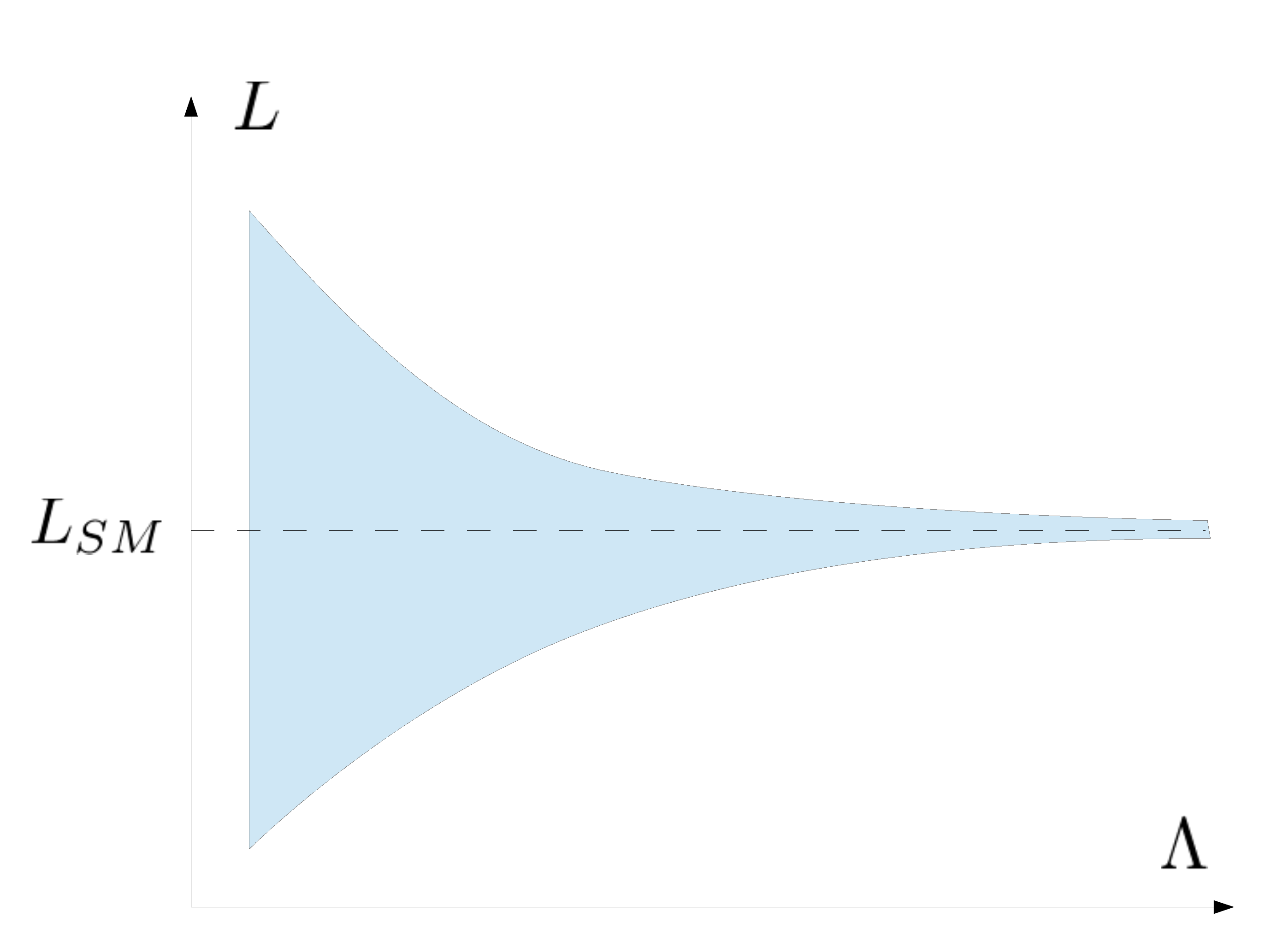}
	\caption{ A picture of the support of the posterior distribution in the $(\Lambda,L)$ plane, $p(\Lambda, L|d,\mathcal{M})$. The support tends to the point $L=L_{SM}$ for $\Lambda\rightarrow \infty$. 
	 \label{fig:pLLambda_pic}}
\end{figure}

Let us first show that the integral of the posterior distribution diverges, \ie
\be
\int_{\,\Lambda,\alpha_i\in\mathcal{D}} \,p(\Lambda,\alpha_i|d,\mathcal{M}) d\mu = \infty \,. \label{eq:post_div}
\ee
To do so, let us rewrite Eq. \eqref{eq:post_div} to make appear the manifold $\Omega_{\Lambda,\hat L}$ defined by fixed values of the likelihood, \footnote{We apply the coarea formula to Eq. \eqref{eq:post_div},
in which we identify the surjective mapping with the likelihood function $L:\mathcal{D}\rightarrow \mathbb{R}$. }
\be
\int_{(\Lambda,L)} d\mu \, \int_{\Omega_{\Lambda, L}} \,
p(\Lambda,\alpha_i|d,\mathcal{M})  \frac{1}{J} \, d \mu(\Omega_{\Lambda, L})
 \,,
 \label{eq:post_div_L}
\ee
where
\be \Omega_{\Lambda,\hat L}=\{\alpha_i| L(\Lambda,\alpha_i)= \hat L \}\,. \label{eq:Omega} \ee
The Jacobian is $ J=(\sum_i (\partial L / \partial \alpha_i)^2 )^{1/2}$.
The marginal posterior in the $(\Lambda,L)$ plane
\be
p(\Lambda,L|d,\mathcal{M})= \int_{\Omega_{\Lambda, L}} \, p(\Lambda,\alpha_i|d,\mathcal{M})  \frac{1}{J} \, d\mu
\ee generates a measure $\nu$ over $(\Lambda,L)$ such that
\be
d\nu= p(\Lambda, L|d,\mathcal{M})d\mu\,
\ee
where $\mu$ is the Lebesgue measure. It is shown in App. \ref{se:asymptotics} that $p(\Lambda, L|d,\mathcal{M})$ tends to a Dirac peak (\ie~$\nu$ tends to the Dirac measure) in the decoupling limit, 
\be
p(\infty, L^{SM}|d,\mathcal{M})\propto \delta(L-L^{SM})\,. \label{eq:post_dirac}
\ee
A schematic picture of the    $p(\Lambda, L|d,\mathcal{M})$ distribution is shown in Fig. \ref{fig:pLLambda_pic}

 Employing Radon-Nikodim (RN) decomposition, the measure $\nu$ can be decomposed as 
\be
\nu=\nu_c+\nu_d\,, \label{eq:RN_dec}
\ee
where $\nu_c$ is absolutely continuous with respect to Lebesgue measure while $\nu_d$ is discrete. 
The discrete measure satisfies
\be
 p(\Lambda, L|L= L^{SM},d,\mathcal{M}) =\frac{d\nu_d}{d\mu}  \,, \label{eq:nu_s}
\ee
and we can then identify our ``excised'' marginal posterior  as
\be
 p^*(\Lambda|d,\mathcal{M})= p(\Lambda, L|L \neq L^{SM},d,\mathcal{M}) =\left|\frac{d\nu_c}{d\mu} \right| \,. \label{eq:nu_c}
\ee
The presence of the absolute value is related to a non-trivial subtlety in the definition of the excised probability space, that is
 discussed in App. \ref{app:sign_flip}. 
In the decomposition of $\nu$ defined by Eqs. \eqref{eq:RN_dec}-\eqref{eq:nu_c} , it turns out that the contribution from the discrete measure $\nu_d$ is infinite, 
\be 
\int_{(\Lambda,L)}\, d\nu_d=\int\, p(\Lambda, L|L= L^{SM},d,\mathcal{M}) d\mu \sim \int \, d\Lambda \, dL\, \Lambda^{-1}\delta(L-L^{SM})=\infty  \,.  \label{eq:int_nu_s}
\ee 
In contrast, one can show that the $\nu_c$ measure leads to a contribution
\be
\int_{(\Lambda,L)}\, d\nu_c \propto \int\, p(\Lambda, L|L\neq L^{SM},d,\mathcal{M}) d\mu\propto \int \,d\Lambda\,\Lambda^{-m-1} \,,  \label{eq:int_nu_c}
\ee
which is finite, for any HDO with dimension $m+4$. 
The proofs of Eqs. \eqref{eq:int_nu_s}, \eqref{eq:int_nu_c}  are given in Appendix \ref{se:integration}.  

From this point of view, it appears that the divergent part of the posterior is localized on the subspace 
$\Omega_{\Lambda,L^{SM}}$. It is precisely the domain where the new physics cannot be tested by the data. 
Requiring testability, we reduce the parameter space to $\mathcal{D}\backslash \Omega_{\Lambda,L^{SM}}$, such that only the contribution Eq. \eqref{eq:int_nu_c} remains in the posterior integral. This contribution being finite, the posterior of testable NP is well proper.

We can check that the requirement of testability is harmless regarding the experimental information. 
Let us recall that the likelihood function comes initially from an experimental probability density function (PDF)  $p_X(x)$ associated with some  observable $X$. We assume that $p_X$ has no discrete component.
The repartition function of the observable $X$ is
\be p(X<x)=\int_{[-\infty,x] }p_X\,d\mu\label{eq:repart}\,.\ee
Expressing $x$ as a function of $(\Lambda,\alpha_i) $, the likelihood function is then $L(\Lambda,\alpha_i)=p_X(x(\Lambda,\alpha_i) )$. The domain $\Omega_{\Lambda,L^{SM}}$ is mapped onto the SM value of the observable $x^{SM}$.  Excluding this domain amounts to excluding the point $x^{SM}$ from the experimental density. A single point having measure zero, this leaves the repartition function unchanged. We conclude that the restriction from $\mathcal{D}$ to $\mathcal{D}\backslash \Omega_{\Lambda,L^{SM}}$  leaves the experimental information  invariant.

\section{The MCMC setup} \label{se:MCMC}

In the present work we are going to evaluate  posterior distributions by means of a Markov Chain Monte Carlo (MCMC) method. The basic idea of a MCMC is setting a random walk in the parameter space such that the density of points asymptotically reproduces a target function, in our case the posterior distribution. Any marginalisation is then performed through a simple binning of the points of the Markov chain along the appropriate dimension. We refer to~\cite{Allanach:2005kz, Trotta:2008qt} for details on MCMCs and Bayesian inference. Our MCMC method uses the Metropolis-Hastings algorithm with a symmetric, Gaussian proposal function.
We check the convergence of our chains using an improved Gelman and Rubin test with multiple chains~\cite{Gelman92}. The first $10^4$ iterations are discarded (burn-in).

Some precautions about the MCMC method are necessary regarding the subtleties about improper posteriors discussed in Sects.  \ref{se:EFT_inference}, \ref{se:consistent_framework}. Indeed, using the MCMC method, we are not working with the exact continuous posterior distributions, as the one discussed in Sect. \ref{se:consistent_framework}. Instead, we are manipulating histograms which are estimators of the exact posteriors. These estimators are discrete distributions
\be
 \hat p_{N,\Delta^{(n)}}(\Lambda,\alpha_i|d,\mathcal{M}) \,,
\ee
where $N$ is the number of points and $\Delta^{(n)}$ is the bin size along the various dimensions. The estimator tends to its  estimand $p(\Lambda,\alpha_i|d,\mathcal{M})$ when $N\rightarrow \infty $, $\Delta \rightarrow 0 $, i.e. in the continuum limit with infinite sampling. 

Notice the bin size can be optimized for a given $N$. Too large bins give a poor estimation of the distribution, while too thin bins suffer from large binomial noise. It exists therefore an optimal bin size to minimize estimators uncertainty. As far as we know it is commonly determined in a ad-hoc way. We proceed similarly in this note.

In the continuous case, we found in Sect. \ref{se:consistent_framework} that the $L=L^{SM}$ subdomain (i.e $\Omega_{\Lambda,L^{SM}}$) shall be excluded to obtain a proper posterior. This feature is translated into the discrete estimator case as follows.
Let us evaluate  $\hat p$  without the $L\neq L^{SM}$ restriction. 
Considering  $\hat p$ in the $(\Lambda,L)$ plane, for $\Lambda\rightarrow \infty$, the only non zero bin of $ \hat p$ is the bin containing the value $L^{SM}$. This is the discrete equivalent of the Dirac peak obtained in Eq. \eqref{eq:post_dirac}.
To obtain the estimator of $p(\Lambda|L\neq L^{SM},d,\mathcal{M})$, we have therefore to excise this bin. This is the discrete equivalent of the  $L\neq L^{SM}$ restriction.
The fact that we exclude a seemingly finite slice of the parameter space should not be surprising, as for the estimator $\hat p$, space is not continuous but discrete. 
Finally, the upper bound $\Lambda<\Lambda_{max}$ also has to be finite in practice. For a given finite $N$ and a given bin size, there exists a finite value   $\Lambda=\tilde \Lambda$ above which all points of $\hat p$ are in the 
$L^{SM}$ bin. In practice one has therefore to make sure that $\Lambda_{max}$ is large enough such that $\tilde \Lambda<\Lambda_{max}$.

\section{Probing $\Lambda$ in the Higgs sector} \label{se:Higgs}

In this Section we apply the inference process defined through Sect. \ref{se:EFT_inference}--\ref{se:MCMC} to the Standard Model Higgs sector extended with higher-dimensional operators. 
The theoretical treatment of HDOs and the analysis of data we used are the same as realized in the recent work \cite{Dumont:2013wma}. Here we briefly review the main points of the analysis, and refer to this work for any further theoretical and experimental  details. 

The Higgs sector is supplemented by a set of CP-even dimension-6 operators, whose basis is chosen to be \footnote{The operator $\mathcal O_6$ plays no role in what follows and is listed here only for completeness.}
\be
\O_6=|H|^6\,,\qquad \O_{D^2}=|H|^2|D_\mu H|^2\,,\qquad \O'_{D^2}=|H^\dagger D_\mu H|^2\,, \label{HDO_O_6}
\ee
\be
\O_{\WW}=H^\dagger H\,(W_{\mu\nu}^a)^2\,,\qquad \label{HDO_O_WW}
\O_{\BB}=H^\dagger H\,(B_{\mu\nu})^2\,,\qquad
\O_{\WB}=H^\dagger\,W_{\mu\nu} H\,B_{\mu\nu}\,,
\ee
\be
\O_{GG}=H^\dagger H\,(G_{\mu\nu}^a)^2\,, \label{HDO_O_GG}
\ee
\be
\O_D=J^a_{H\,\mu}\,J^a_{\mu}\,,\qquad
\O_{D}'=J^Y_{H\,\mu}\,J^Y_{\mu}\,,
\ee
\be
\O_f= 2 y_f\,|H|^2 \, H\bar f_L f_R \,. \label{HDO_O_f}
\ee
Here $J_H$ and $J_{f}$ are $SU(2)$ or $U(1)_Y$ currents involving the Higgs field and the fermion $f$  respectively, and $J=\sum_f J_{f}$ are the SM fermion currents coupled to $B_\mu$ and $W_\mu$.

This choice of basis is such that the field strength--Higgs operators $\mathcal{O}_{FF}$'s cannot be generated at tree-level in a perturbative UV theory. We therefore consider two general cases, ``democratic HDOs'' and ``loop-suppressed $\mathcal O_{FF}$'s\,'', depending on whether or not the $\mathcal O_{FF}$'s are loop-suppressed with respect to the other HDOs. 
Moreover, in important classes of models like for the R-parity conserving MSSM, the HDOs can only be generated at one-loop. We will therefore consider two cases within the democratic HDOs scenario, one with tree-level HDOs,  $\alpha_i\in[-16\pi^2,16\pi^2]$, and one with loop-level HDOs,  $\alpha_i\in[-1,1]$. 
For the case of loop-suppressed $\mathcal O_{FF}$'s, we assume that the unsuppressed HDOs are generated at tree-level. We therefore investigate three scenarios whose features are summarized in  Tab. \ref{tab_scenarios}. In case of tree-level HDOs, perturbativity of the HDO expansion $|\alpha|/\Lambda^2<1/v^2$ imposes an additional constraint for $\Lambda<4\pi v$.
We take custodial symmetry to be an exact symmetry of the theory, such that the operators $\O_D'$, $\O_{D^2}'$ are set to zero. 
Finally, we emphasize that these scenarios are generic, in the sense that they encompass all known UV models in addition to the realizations not yet thought of. This implies that features predicted only by specific UV models, like suppression of HDOs or precise cancellations between HDOs, will get a small statistical weight, as we consider the whole set of UV realizations.



\begin{table}
\center
\begin{tabular}{|c|c|c|c|}
\hline
& \multicolumn{2}{|c|}{ Democratic HDOs} &  Loop-suppressed $\mathcal O_{FF}$'s \\
\hline
 & Tree-level & One-loop  & \\
 \hline
$\Lambda$   &  $[v, \infty[$ & $[v, \infty[$ &  $[v, \infty[$ \\
\hline
$|\alpha_{FF}|$   & $ [0,\Lambda^2/v^2]$ if $\Lambda<4\pi v$  & $ [0,1]$  & $ [0,1]$   \\
                &  $ [0,16\pi^2]$ else, &   &  \\
\hline
Other $|\alpha|$   & $ [0,\Lambda^2/v^2]$ if $\Lambda<4\pi v$  & $ [0,1]$  & $ [0,\Lambda^2/v^2]$ if $\Lambda<4\pi v$  \\
                &  $ [0,16\pi^2]$ else, &  &$  [0,16\pi^2]$ else. \\
\hline
\end{tabular}
\caption{Summary of the setup of the scan in the three scenarios we consider. The $\alpha_{FF}$ coefficients (where $FF = \WW,\,WB,\,BB,\,GG$) correspond to the field-strength--Higgs operators. In both cases we take custodial symmetry to be an unbroken symmetry.  } \label{tab_scenarios}
\end{table}

Concerning data, we take into account the results from Higgs searches at the LHC and at Tevatron as well as electroweak precision observables and trilinear gauge couplings. 
Higgs results \cite{tevatron:2012zzl,ATLAS-CONF-2013-012, ATLAS-CONF-2013-014, ATLAS-CONF-2013-013,ATLAS-CONF-2013-030,ATLAS-CONF-2013-034,ATLAS-CONF-2012-161,CMS-PAS-HIG-13-001,CMS-PAS-HIG-13-002,CMS-PAS-HIG-13-003,CMS-PAS-HIG-12-039,CMS-PAS-HIG-12-042,CMS-PAS-HIG-12-045,CMS-PAS-HIG-12-044,CMS-PAS-HIG-12-025,CMS-PAS-HIG-13-004,CMS-PAS-HIG-13-006,HCPHiggsTevatron,HCPtevBBtalk} have to be exploited with care as HDOs modify both Higgs decays and production. We use results (partly) accounting for  correlations between the subchannels when they are available.
When estimated decomposition into production channels are unavailable, we take the relative ratios of production cross sections for a SM Higgs~\cite{HXSWG, tevatron:2012zzl}  as a reasonable approximation.
The Higgs mass is set to $m_h = 125.5$~GeV, close to the combined mass measurement from the two experiments, since it is not yet possible to take it as a nuisance parameter without losing the correlations between production channels. 
We take into account the electroweak precision observables using the Peskin-Takeuchi $S$ and $T$ parameters~\cite{Peskin:1990zt,Peskin:1991sw}. 
Beyond $S$ and $T$, the $W$ and $Y$ parameters~\cite{Barbieri:2004qk} should be used in the HDO framework. However the constraints arising from these parameters are by far negligible with respect to our other constraints.  Experimental values of $S$ and $T$ are taken from the latest SM fit~\cite{Baak:2012kk}, $S=0.05 \pm 0.09$ and $T= 0.08 \pm 0.07$ with a correlation coefficient of $0.91$.
Regarding constraints on TGV, we take into account the LEP measurements~\cite{TGV}.
%

\begin{figure}
\begin{picture}(400,100)
\put(0,0){		\includegraphics[trim=0cm 7.5cm 0cm 7cm, clip=true,width=7cm]{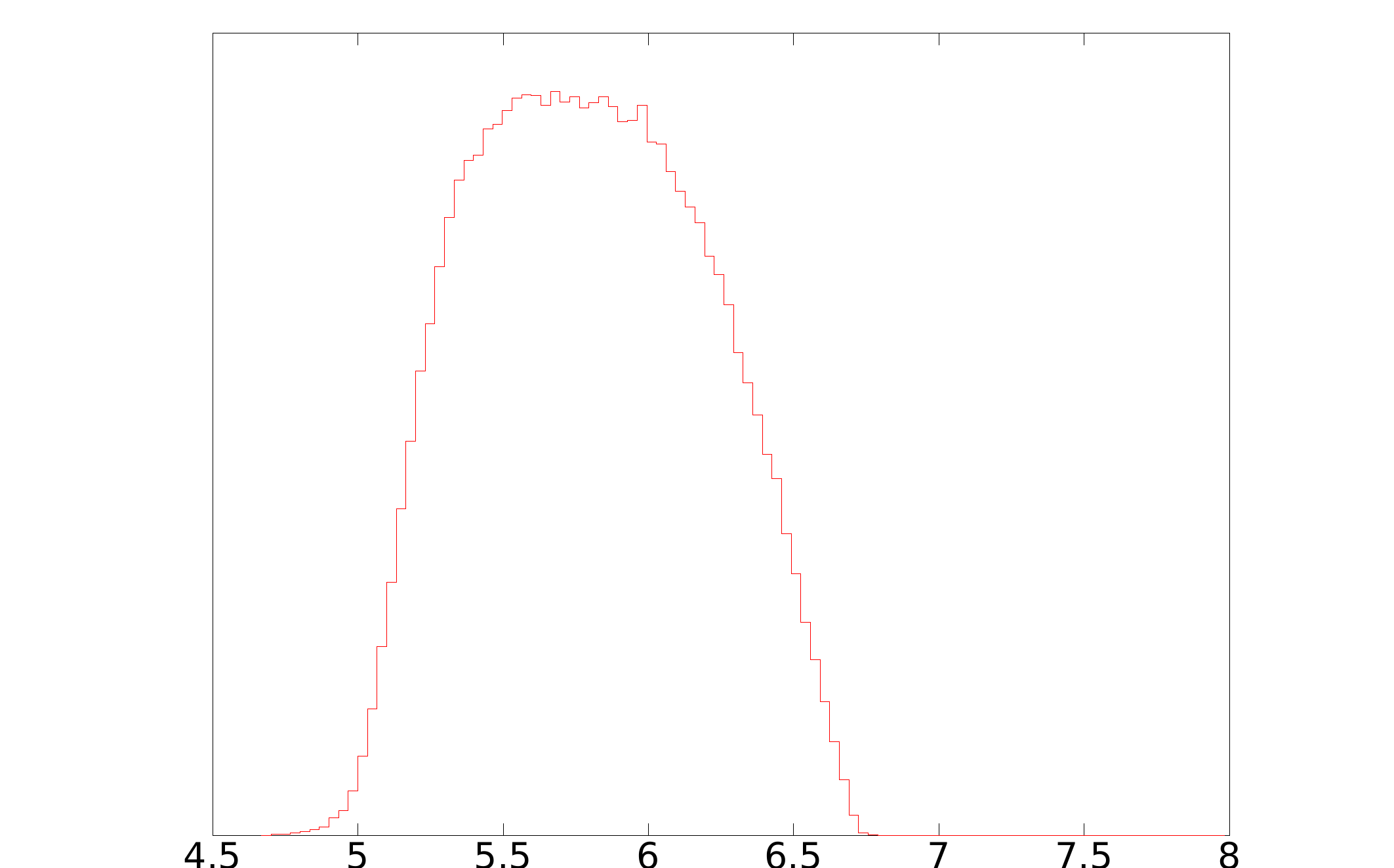}}
\put(200,0){		\includegraphics[trim=0cm 7.5cm 0cm 7cm, clip=true,width=7cm]{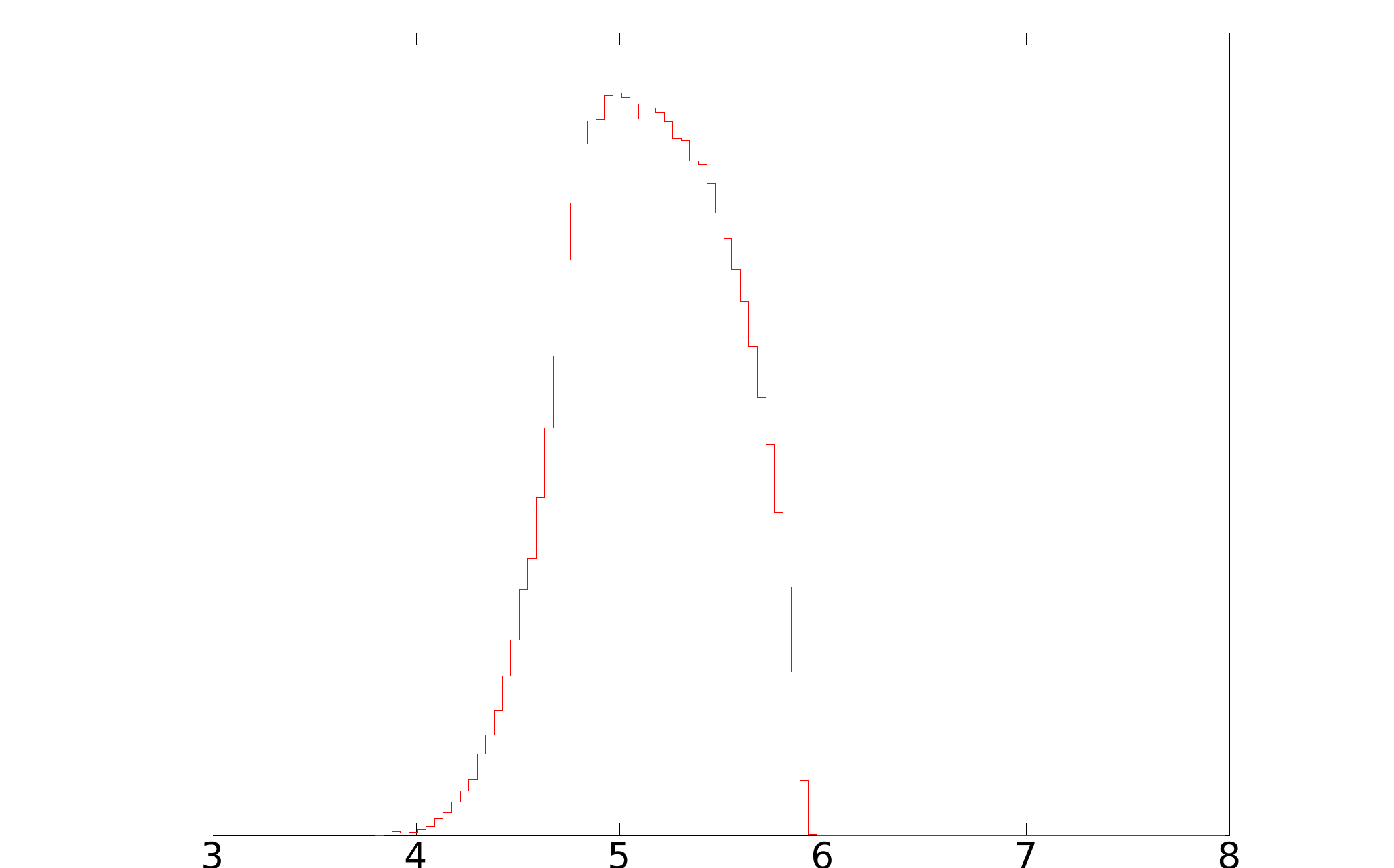}}
\put(70,-10){$\small{\log_{10}(\Lambda/\textrm{GeV})}$}
\put(270,-10){$\small{\log_{10}(\Lambda/\textrm{GeV})}$}
\end{picture}
\vspace{0.5cm}
	\caption{Posterior PDFs of $\Lambda$ ( $p^*(\Lambda|d,\mathcal{M})$ ) in GeV units for tree-level democratic HDOs (right) and loop-suppressed $\O_{FF}$'s (left). The PDF of $\Lambda$ for one-loop democratic HDOs has the same shape as for tree-level democratic HDOs and is shifted by $\log_{10} 4\pi$ towards low values of $\Lambda$.
	 \label{fig:pLambda}}
\end{figure}

Applying the method described in  Sect. \ref{se:EFT_inference}--\ref{se:MCMC}, we obtain the normalizable posterior distributions $p^*(\Lambda|d)$. One can always normalize them to unity such that  we will designate them as probability density functions (PDFs). It turns out that the posterior PDF of the NP scale for tree-level and one-loop democratic HDO is essentially the same under a shift $\log_{10}\Lambda\rightarrow \log_{10}\Lambda -\log_{10}(4\pi)\approx \log_{10}\Lambda- 1.10$, \ie~a rescaling $\Lambda\rightarrow \Lambda/4\pi$.
This happens because the region $|\alpha|\in[0,\Lambda^2/v^2]$ for tree-level HDOs has a negligible impact on the posterior, such that the tree-level and one-loop scenarios can be identified through a rescaling of $\Lambda$. The posterior PDFs of the NP scale for the various scenarios are shown in Fig. \ref{fig:pLambda}.
 The $68\%$ and $95\%$  Bayesian credible intervals (BCIs) of $\Lambda$ for democratic HDOs are respectively 
 $[200,1400]$, $[123,3300]$ TeV for tree-level HDOs and $[16,110]$, $[9.8,260]$ TeV for loop-level HDOs.
We find $68\%$ and $95\%$ BCIs of $[62,533]$, $[28,1200]$ TeV for the scenario of  loop-suppressed $\O_{FF}$'s.


  We find the leading constraint on $\Lambda$ to be the Higgs data for democratic HDOs, while these are the electroweak observables for loop-suppressed $\O_{FF}$'s. 
 This can be understood as follows. The $\O_{FF}$ operators are mapped onto field strength--Higgs anomalous couplings, among which the $\zeta_g h (G_{\mu\nu})^2$ and $\zeta_\gamma h (F_{\mu\nu})^2$ couplings. Given that the corresponding SM couplings are generated at one-loop, $\zeta_{g,\gamma}$ need to be sensibly suppressed to not induce large deviations in the predictions of  gluon fusion and $h\rightarrow \gamma\gamma$ processes (see \cite{Dumont:2013wma} for  details).   
 For democratic HDOs, this need of small $\zeta_{g,\gamma}$ pushes $\Lambda$ to high values in order to suppress the $\O_{FF}$'s. In contrast, for the scenario of  loop-suppressed $\O_{FF}$'s, the $\zeta_{g,\gamma}$'s are already loop-suppressed with respect to other anomalous couplings by assumption. This alleviates the aforementioned constraint, leaving the $S$, $T$ measurements as leading constraints.

Having identified the leading constraints, we may comment about the necessary conditions allowing more specific UV models to reach lower values of $\Lambda$. For models having democratic HDOs, a suppressed $\O_{GG}$ is required to reduce the $\zeta_g$ coupling.  
The $\zeta_\gamma$ coupling being proportional to $s^2_w\alpha_{WW}+ c^2_w \alpha_{BB}-\frac{1}{2}c_ws_w \alpha_{WB}$, precise cancellations among these various terms may occur within an appropriate UV model, while they are unprobable (\ie~fine-tuned) in the generic scenario.  Note both conditions on $\zeta_g$ and $\zeta_\gamma$ need to be fulfilled in order to lower the values of $\Lambda$. If only one of the $\zeta$'s is suppressed, the outcome will still remain similar to the 
left plot of Fig. \ref{fig:pLambda}. This occurs in particular when these $\zeta$'s are generated perturbatively. In that case  one has $\zeta_g/\zeta_\gamma\approx g_s^2/g_Y^2\gg1$, such that $\zeta_\gamma$ is naturally suppressed with respect to $\zeta_g$, which then becomes the 
 leading constraint.
Concerning models with loop-suppressed $\O_{FF}$'s, the main condition to reach a lower $\Lambda$ is to have a suppressed $\O_D$. This operator induces the main contribution to the $S$ parameter, $\alpha S \approx s^2_w \alpha_D v^2/\Lambda^2$,  other contributions to $S,T$ being loop-suppressed (see \cite{Dumont:2013wma}). 


%

The PDFs presented above are given for an optimal size of the bins. \footnote{Namely we use $400$ bins for a sampling $N=O(2\times10^6)$.} To exemplify the uncertainty associated with the MCMC estimation of the PDFs, we compute the $95\%$ BCIs obtained with twice more bins and twice less bins. We find the variations over $\log_{10}\Lambda$ to be $O(2\%)$. 
The origin of these variations lies in the uncertainty inherent to the concrete estimation method presented in Sect. \ref{se:MCMC}, and is not related to the formal inference process described in Sects. \ref{se:EFT_inference}, \ref{se:consistent_framework}.
 


\section{Conclusion} \label{se:conclusion}

Whenever one considers a set of data -- showing or not a significant deviation from the Standard Model, 
it is interesting to ask  what information can be obtained about the energy scale of a possible underlying new physics.  
We present a method to consistently infer the distribution of $\Lambda$ from any dataset. In doing so we use a statistical view of the unknown NP  parametrized by higher-dimensional operators.
To obtain a proper posterior, necessary to create Bayesian credible intervals, we point out the requirement that NP has to be testable by the data. 

We formally demonstrate using Lebesgue integration that this requirement implies  proper posteriors. In doing so we introduce a subspace where the likelihood itself is taken as a random variable. Some conceptual subtleties related to this trick are discussed, and a helpful toy model  is introduced in the appendix.
Given that Monte Carlo Markov Chains methods are commonly used to realize statistical inference, we describe the concrete implementation of this inference process  in MCMCs.

As an illustration, we apply our approach to the SM Higgs sector, in light of recent data. Building on the recent work \cite{Dumont:2013wma}, we consider the scenarios of democratic HDOs and loop-suppressed $\O_{FF}$'s.  For democratic HDOs, we obtain $95\%$ Bayesian credible intervals of $[123,3300]$ TeV  and $[9.8,260]$ TeV, for respectively tree-level and loop-generated HDOs. For loop-suppressed $\O_{FF}$'s, we find the $95\%$ BCI  $[28,1200]$ TeV, assuming that unsuppressed HDOs are generated at tree-level.
More specific UV models suppressing some particular HDOs or  predicting otherwise fine-tuned  relations are necessary to favour lower values of the NP scale.


\section*{Acknowledgements} 

The author would like to thank Gero von Gersdorff and B\'eranger Dumont for intense discussions relative to this study. The author is grateful to Ivan Gordeli and Gero von Gersdorff for reading the manuscript. The author acknowledges the Brazilian Ministry of Science, Technology and Innovation for financial support, and the LPSC and the Ecole Polytechnique for hospitality during part of this work.

\vspace*{2cm}

\appendix

\noindent{\Large\bf Appendix}

\vspace*{0.5cm}

The effective Lagrangian Eq. \eqref{L_eff} contains in general HDOs of arbitrary dimensions. Here, for simplicity we will consider HDOs of a single dimension labelled by $m$. The generalization to the case of HDOs with arbitrary dimension is straightforward.

\section{Asymptotics of $p(\Lambda,L|d,\mathcal{M})$} \label{se:asymptotics}

Let us show that the marginal posterior in the $(\Lambda,L)$ plane
\be
p(\Lambda,L|d,\mathcal{M})= \int_{\Omega_{\Lambda, L}}\, p(\Lambda,\alpha_i|d,\mathcal{M})  \frac{1}{J} \, d\mu 
\ee
tends to be proportional to the Dirac peak in the decoupling limit,
\be
p(\Lambda=\infty,L)\propto  \delta(L-L^{SM})\,. \label{eq:delta}
\ee

\textit{Proof}
In the decoupling limit $\Lambda \rightarrow \infty $, $L$ tends to $L^{SM}$. 
Thus, for any arbitrary small $\delta L>0$, it exists a finite $\Lambda=\tilde \Lambda$ such that  $\Omega_{\Lambda, L}=\emptyset$ for any $\Lambda>\tilde \Lambda$ and $|L-L^{SM}|>\delta L$.
In the decoupling limit with  $L\neq L^{SM} $, the integration domain
 $\Omega_{\Lambda>\tilde \Lambda, L}$ therefore reduces to the null set. This implies
 \be p(\Lambda\rightarrow \infty, L|L\neq L^{SM}  d,\mathcal{M})=0 \,. \label{eq:LneqLSM}\ee
 Let us now study the behaviour for $L= L^{SM}$. 
Defining  $\partial_i L= \partial L(x_i)/\partial x_i$,  a $\Lambda^{-m}$ factor out from the Jacobian $J=(\sum_i(\partial L/\partial \alpha_i )^2)^{1/2}$,  
\be J=\Lambda^{-m}(\sum_i (\partial_i L)^2 )^{1/2}\,.\ee 
The $\partial_i L$ are finite by hypothesis, such that $J=O(\Lambda^{-m}) $. The posterior  $p(\Lambda,\alpha_i|d,\mathcal{M})$ is therefore  $O(\Lambda^{m-1})$ once one takes into account the log prior. For any  $m \geq 2$, the distribution gets therefore infinite,
\be
 p(\Lambda=\infty, L|L= L^{SM},d,\mathcal{M})= \infty\,. \label{eq:L=LSM}
\ee
We deduce from Eqs. \eqref{eq:LneqLSM}, \eqref{eq:L=LSM} that $ p(\Lambda=\infty, L|L= L^{SM},d,\mathcal{M})$ is proportional to a Dirac peak centred on $L=L^{SM}$. $\square$

\section{Integration of  the posterior } \label{se:integration}

Starting from 
\be
d\nu_d=p(\Lambda, L|L\neq L^{SM},d,\mathcal{M}) d\mu \,,
\ee
\be
d|\nu_c|=p(\Lambda, L|L= L^{SM},d,\mathcal{M}) d\mu \,,
\ee
we want to show that $\int_{(\Lambda,L)} d\nu_d=\infty$ and that $\int_{(\Lambda,L)} d\nu_c$ is finite. One assumes $\nu_c>0$.

\textit{Proof}
Let us denote the marginal posterior along $\Lambda$  as
\be
 p(\Lambda,d,\mathcal{M}) \equiv f(\Lambda)
\ee
for simplicity. We write
 $f=f^*+f^{SM}$ with
\be f^*(\Lambda)=\int\, p(\Lambda, L|L\neq L^{SM},d,\mathcal{M})\, d\mu(L) \ee and 
\be f^{SM}(\Lambda)=\int \, p(\Lambda, L|L= L^{SM},d,\mathcal{M})\,d\mu(L)\,, \ee
such that
$\int d\nu_d=\int \,f^{SM}(\Lambda) \,d\mu $ and $\int d\nu_c=\int\, f^{*}(\Lambda)\,d\mu $.

We define the simple function
\be
f_n=\sum_{k=0}^{n2^n-1}\frac{k}{2^n}\mathbb{I}(E_k)+n\mathbb{I}(n,\infty)\,.
\ee
with $E_k=[f^{-1}(\frac{k}{2^n}),f^{-1}(\frac{k+1}{2^n})]$.
$f_n$ converges pointwise to $f$ and we have $f_n(x)\leq f(x)$, such that $\int f d \mu=\lim_{n\rightarrow \infty}\int f_n d \mu$ by the Monotone Convergence Theorem (MCT). 
We define $k^{SM}$ such that $k^{SM}/n< f^{SM} <(k^{SM}+1)/n$. 
We then have $f_n=f_n^{SM}+f_n^*$ with
\be
f^{SM}_n=\frac{k^{SM}}{2^n}\mathbb{I}(E_k^{SM})\,,
\ee
\be
f^{*}_n=\sum_{k=0,k\neq k^{SM}}^{n2^n-1}\frac{k}{2^n}\mathbb{I}(E_k)+n\mathbb{I}(n,\infty)\,.
\ee
and $\lim_{n\rightarrow \infty} f_n^{SM}=f^{SM}$, $\lim_{n\rightarrow \infty} f_n^{*}=f^{*}$.

Let us compute $\int f^{SM}_n d \mu$ where $\mu$ is the Lebesgue measure. 
Given that $L\rightarrow L^{SM}$ for $\Lambda\rightarrow \infty$, for any arbitrary small $\delta L=k^{SM}/2^n-L^{SM}$, it exists a finite $\tilde \Lambda$  such that $f\in E_{k^{SM}}$ for any $\Lambda\in [\tilde \Lambda, \infty]$. Therefore $\mu(E_{k^{SM}})=\infty$. This implies $\int \,f_n^{SM}(\Lambda) \,d\mu=\infty$, then  $\int \,f^{SM}(\Lambda) \,d\mu=\infty$ by the MCT, and thus $\int d\nu_d=\infty$.

Let us now compute  $\int f^{*}_n d \mu$.  $\mu(E_{k\neq k^{SM}})$ is finite. We have to show that the sum over $n$ converges. 
To do so we first simplify $f^{*}$ using the $\Lambda\rightarrow \infty$ limit. The Limit Comparison Test (LCT) will ensure that the simpler function has the same integrability features as $f^{*}$. We will denote the successive simplified functions by $\hat f^{*}$.

 We factorize the $\Lambda$ prior and factorize the likelihood function out from the first integral such that
\be
f^*(\Lambda)=\frac{1}{\Lambda} \int\, d\mu(L) \,L\, \int_{\Omega_{\Lambda, L}} \,
 \frac{1}{J} \, d \mu(\Omega_{\Lambda, L})
 \,. \label{eq:F(Lambda)}
 \ee
We replace the derivatives $\partial_i L$ in the Jacobian of Eq. \eqref{eq:F(Lambda)} by their values at $\Lambda\rightarrow \infty$. The LCT ensures that this simplified function as the same integrability properties as Eq. \eqref{eq:F(Lambda)} as their limits for $\Lambda\rightarrow \infty$ are the same. We can then integrate over $d \mu(\Omega_{\Lambda, L})$ and obtain
\be
\hat f^*\propto\Lambda^{m-1} \int\, d\mu(L) \, L    \, \mu(\Omega_{\Lambda, L})
 \,.
 \ee
For any finite $\tilde \Lambda$, one can expand the likelihood with respect to $\tilde \Lambda / \Lambda$, 
\be
L=L^{SM}+ \left. \frac{\partial L}{\partial \Lambda^{-1}} \right|_{\Lambda\rightarrow\infty} . \frac{1}{\Lambda}  +O(\frac{\tilde \Lambda^2}{\Lambda^2}) \,.
\ee
 $L$ can be reexpressed as 
\be
L=L^{SM}+ \left. \partial_i L \right|_{\infty} . \alpha_i \frac{\tilde \Lambda}{\Lambda}  +O(\frac{\tilde \Lambda^2}{\Lambda^2}) \,. \label{eq:hyper}
\ee
The LCT ensures that one can replace $L$ by its truncated expansion to study the integrability of $f^*$.
In this limit, $\mu(\Omega_{\Lambda, L})$ is the volume of a hyperplane  in the $\{\alpha_i\}$ space defined by Eq. \eqref{eq:hyper}. We can write it as $\mu(\Omega_{\Lambda, L})=\mathcal{V}\{(L-L^{SM})^2\Lambda^{2m}\}$ such that the squared likelihood and the $\Lambda$ dependence appear explicitely. We are left with studying the integrability of 
\be
\hat f^*=\Lambda^{m-1} \int\, d\mu(L) \, L    \, \mathcal{V}\{(L-L^{SM})^2\Lambda^{2m}\}
 \,.
 \ee
Changing variable $(L-L^{SM})^2\Lambda^{2m}\rightarrow (L-L^{SM})^2 $ factors out a $\Lambda^{-2m}$ factor. The remaining integral $\frac{1}{2}\int\, d(L-L^{SM})^2 \, \mathcal{V}\{(L-L^{SM})^2\}$
gives $\frac{1}{2}\mu(\{\alpha_i\})$ which is bounded by hypothesis. \footnote{Recall that this is imposed by perturbativity of couplings of the UV theory.} For example in the tree-level democratic HDOs case we have $\mu(\{\alpha_i\})=(32\pi^2)^m$.
We have therefore $\hat f^*= \frac{1}{2}\Lambda^{-m-1}\mu(\{\alpha_i\})$. $\hat f^*$ being Riemann integrable  over $[\Lambda_{min},\infty]$ and absolutely convergent, it is therefore Lebesgue integrable. We deduce that $\int \hat f^*_n d\mu$ converges for $n\rightarrow \infty$ , thus $\int f^*_n d \mu$ converges as well by the LCT, the integral $\int f^* d\mu$ is therefore finite and so is  $\int d\nu_c$.
$\square$

\section{Probability definition in the excised space}
\label{app:sign_flip}

Here we discuss the subtlety  that leads to the apparition of the  absolute value on $|d\nu_c/d\mu|$ in Eq. \eqref{eq:nu_c}. We stress that this discussion mainly matters at the formal level. In practice, for example when computing the   posterior $p^*(\Lambda|d)$ using the MCMC method of Sec. \ref{se:MCMC}, this question will not appear.

First,  notice that we expressed our posterior distribution as  a function  of the likelihood $L$. This is perfectly allowed, as the 
likelihood can be just seen as a random variable as another.  However the likelihood is also a conditional probability. Our ``excised''  space $\mathcal{D}\backslash \Omega_{\Lambda,L^{SM}}$ is thus rather particular. 

Second, let us note that in  the Kolmogorov axioms of probability, the    positivity axiom can be seen  as a simple  sign convention. For any   sample space $\Omega$, requiring $p(\Omega)=-1$ and  $p(E)\leq 0$, $\forall E\in\Omega$,  the subsequent  results just change by a sign flip. Let us denote by $p^{(-)}$ the probabilities defined in this way, and by $p^{(+)}$ the usual positive probabilities. 
One of the  consequence of using the $p^{(-)}$ system 
 is that  the expectation of a random variable $X$ is given by \be\left<X\right>=-\int dx \,x\, p^{(-)}_X(x) \,.\ee
When using such convention, a crucial point is that the conditional probabilities must still be taken positive, contrary to the actual probabilities -- inconsistencies would appear otherwise due to the probability  multiplications. 
The freedom to switch between  the $p^{(+)}$ and $p^{(-)}$ system of probabilities is just a symmetry of the classical probability theory.

Keeping these points in mind, let us now compute the naive expectation $\left<L^*\right>_\Lambda'$  of the likelihood $L$  over the excised parameter space $\mathcal{D}\backslash \Omega_{\Lambda,L^{SM}}$.  To do so we use  the RN decomposition of Eq. \eqref{eq:RN_dec}, and obtain
\be
\left<L^*\right>_\Lambda'= \int \, L\,  d \nu_c = \left<L\right>_\Lambda -L^{SM}\,.
\ee
It is clear that $\left<L\right>_\Lambda$ is not necessarily larger than $L^{SM}$. This typically happens when data disfavor the model with respect to the SM. We deduce that  $\left<L^*\right>_\Lambda'$ can take both signs. But  in the two paragraphs above, we emphasized 
that $L$ is a conditional probability, and as such must be positive whatsoever. We conclude that $\nu_c$ has to be taken as a probability measure of the   $p^{(-)}$ kind, whenever $\left<L\right>_\Lambda -L^{SM}<0$. The actual expectation is then  $\left<L^*\right>_\Lambda=-\int \, L\, d\nu_c$, which is positive as it should. We thus end up with the prescription that 
the measure $\nu_c$ is taken as a probability $p^{(+)}$ or $p^{(-)}$ when 
 $\left<L\right>_\Lambda -L^{SM}$ is respectively  positive or negative. Finally, as soon as we restrict ourselves to  the excised space, we always have the freedom to switch between $p^{(-)}$ and $p^{(+)}$. Choosing to deal only with $p^{(+)}$, the
 probability density over the excised space is expressed as
\be
p^*(\Lambda|d,\mathcal{M})=
\begin{cases}
d (\nu-\nu_s)/d\mu & \textrm{if}\,\left<L\right>_\Lambda -L^{SM}>0
\\
 d (\nu_s-\nu)/d\mu & \textrm{if}\,\left<L\right>_\Lambda -L^{SM}<0\,,
\end{cases}
\ee 
hence the absolute value in Eq. \eqref{eq:nu_c}.

\section{The BSM coin} \label{app:BSM_coin}

To exemplify our approach, let us adopt a simple NP model.
Suppose that the SM predicts that a certain coin is fair. It comes Heads or Tails with probability $1/2$. Suppose that a HDO modifies the probability such that the coin is not fair anymore,\footnote{We are grateful to the referee for pointing out to us this simple example.}
\be
p(H|\Lambda,\alpha)=1/2+\alpha/\Lambda\,,\,\,\,p(T|\Lambda)=1/2-\alpha/\Lambda\,.
\ee
The SM is recovered for $\Lambda\rightarrow \infty$, or if $\alpha=0$.

$\Lambda$ is given a logarithmic prior over $[2,\infty[$, $\alpha$ is given a flat prior over $[-1,1]$.
Let us assume that the coin is tossed twice and comes down "H,T". 
We toss the coin only twice for simplicity of the subsequent expressions. A more complicated likelihood would unnecessarily complicate the formulas. In doing so, data favor the SM hypothesis. We can thus  expect a likelihood $\left<L\right>_\Lambda<L_{SM}$.

The SM likelihood $L_{SM}$ is \be
p(HT,SM)=(\frac{1}{2})^2=\frac{1}{4}\,.
\ee
We now work out the SM+HDO likelihood, 
 \be
p(HT|\Lambda,\alpha)=(1/2+\alpha/\Lambda)(1/2-\alpha/\Lambda)=\frac{1}{4}-\frac{\alpha^2}{\Lambda^2}\,.
\ee

Let us first compute the posterior PDF of $\Lambda$ without any ''excision". It is given by
\be
p(\Lambda|HT)\propto \int d\alpha p(HT|\Lambda,\alpha)\,p(\Lambda)\, p(\alpha)
\ee
\be
p(\Lambda|HT)\propto \int_{-1}^1 d\alpha \,(\frac{1}{4}-\frac{\alpha^2}{\Lambda^2})\, \frac{1}{\Lambda}\,\frac{1}{2}
\ee
\be
p(\Lambda|HT)\propto (\frac{1}{4\Lambda} -\frac{1}{3\Lambda^3})\label{posterior_simple}
\ee
As expected the posterior $p(\Lambda|HT)$ is  not integrable over $[2,\infty[$, i.e it is  improper. 

Let us now proceed to the excision. What we want to compute is the distribution $p^*(\Lambda|HT )$, 
\be
p^*(\Lambda|HT )=\int_{L\neq L_{SM}}  p(\Lambda,L|HT) \,dL\,. \label{eq:pstar_coin}
\ee
From a one-to-one variable change using $\alpha=\Lambda\sqrt{1/4-L}$,  we compute 
\be
p(\Lambda,L|HT)\propto
  \frac{L}{\sqrt{1- 4L}}\,.
\ee
The measure  
\be
d\nu=p(\Lambda,L|HT) dL
\ee
is singular in $L=1/4=L_{SM}$, such that one can decompose it as $d\nu=d\nu_{d}+d\nu_{c}$ where
\be
d\nu_{d}=p(\Lambda,L|L=L_{SM,}HT)\delta(L-L_{SM})dL\,.
\ee
\be
d|\nu_{c}|=\left. p(\Lambda,L|L\neq L_{SM},HT)\right|dL\,.
\ee
Plugging the decomposition into the integral of Eq. \eqref{eq:pstar_coin}, we have
\be
p^*(\Lambda|HT )=\left|\int p(\Lambda,L|HT) dL- p(\Lambda,L_{SM}|HT) \right| \,.
\ee

Let us work out the two terms on the right-hand side of the equation above. The first one is just Eq. \eqref{posterior_simple} written differently,
\be
\int p(\Lambda,L|HT) dL\propto\int_{\frac{1}{4}-\frac{1}{\Lambda^2}}^{\frac{1}{4}}dL \,\frac{L}{\sqrt{1- 4L}}=-\frac{1}{12}\left[ \sqrt{1-4L}(2L+1)     \right]_{\frac{1}{4}-\frac{1}{\Lambda^2}}^{\frac{1}{4}}=(\frac{1}{4\Lambda} -\frac{1}{3\Lambda^3})
\ee
The second one is 
\be
 p(\Lambda,L=L_{SM}|HT)\propto\frac{L_{SM}}{\Lambda}=\frac{1}{4\Lambda}\,.
\ee
The proportionality constant is the same for both terms.
The divergent piece cancels between both terms, leaving
\be
p^*(\Lambda|HT)\propto\frac{1}{3\Lambda^3}\,.
\ee

\begin{figure}
	\centering
		\includegraphics[trim=0cm 0cm 0cm 0cm, clip=true,width=7cm]{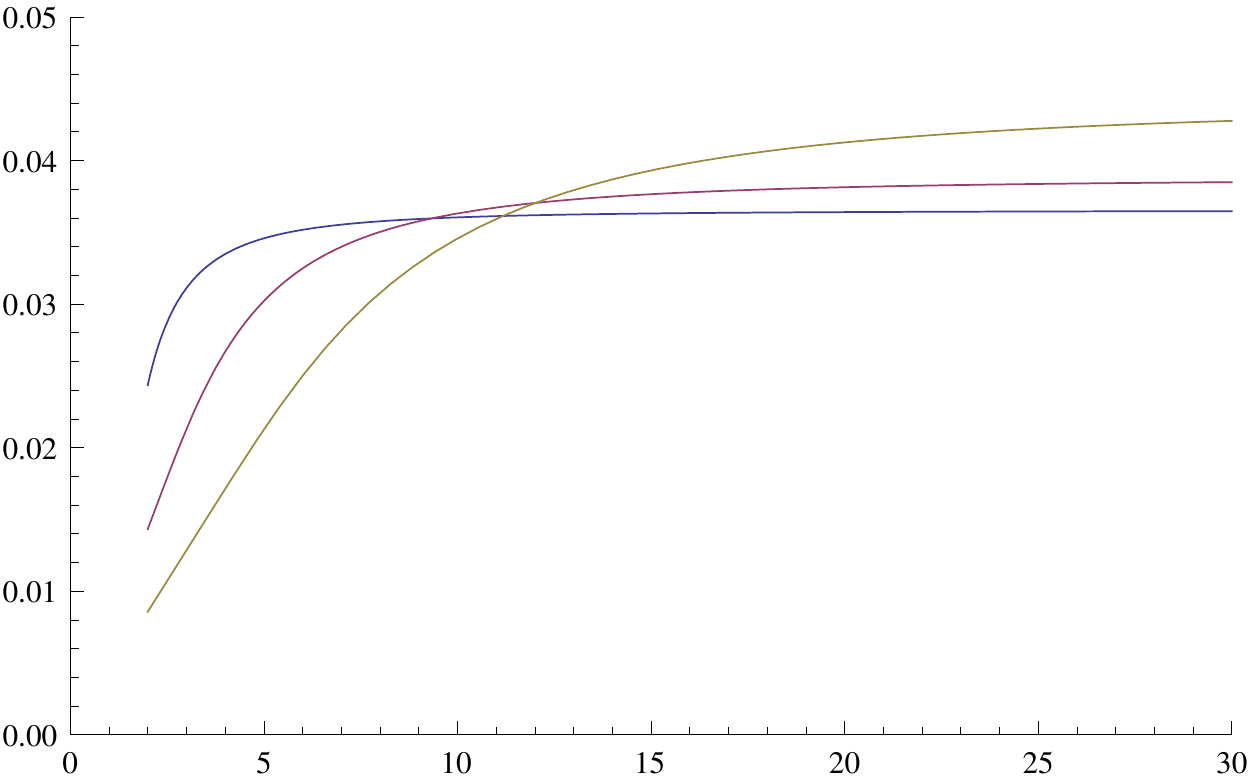}		
		\includegraphics[trim=0cm 0cm 0cm 0cm, clip=true,width=7cm]{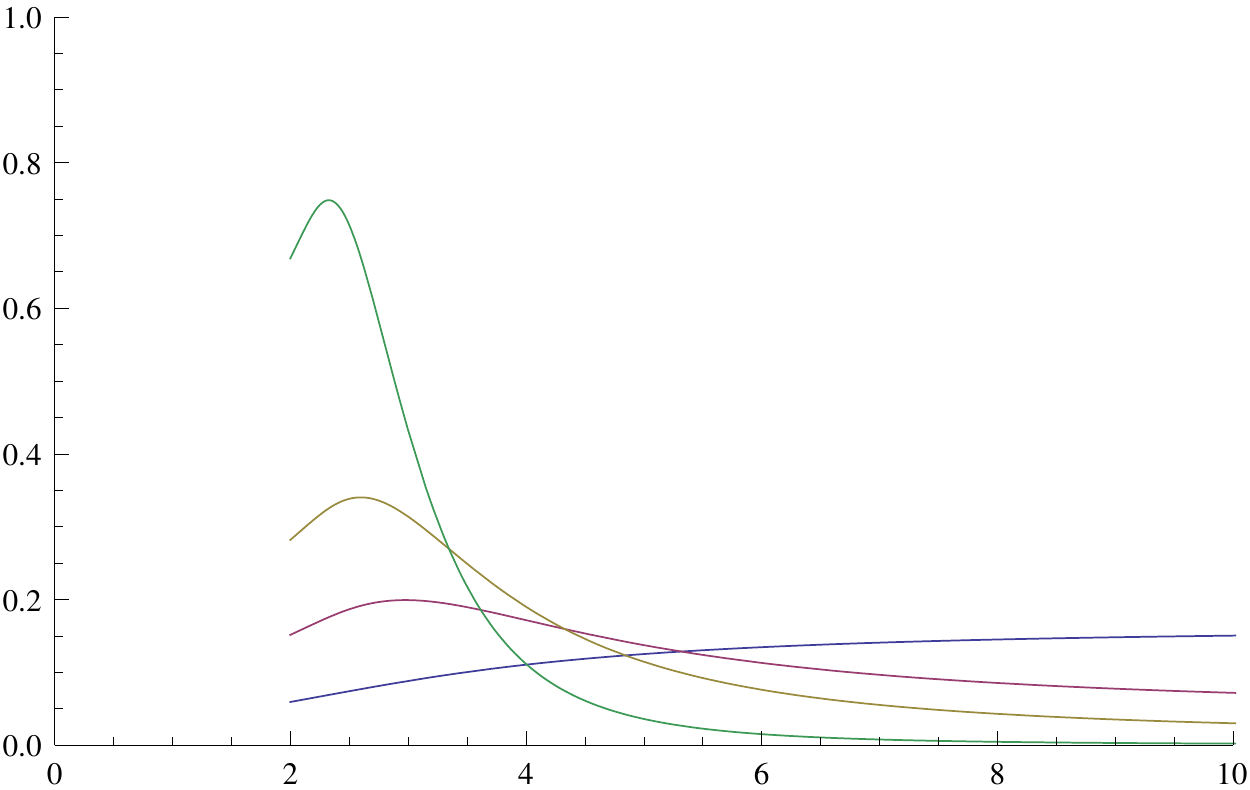}
		\includegraphics[trim=0cm 0cm 0cm 0cm, clip=true,width=7cm]{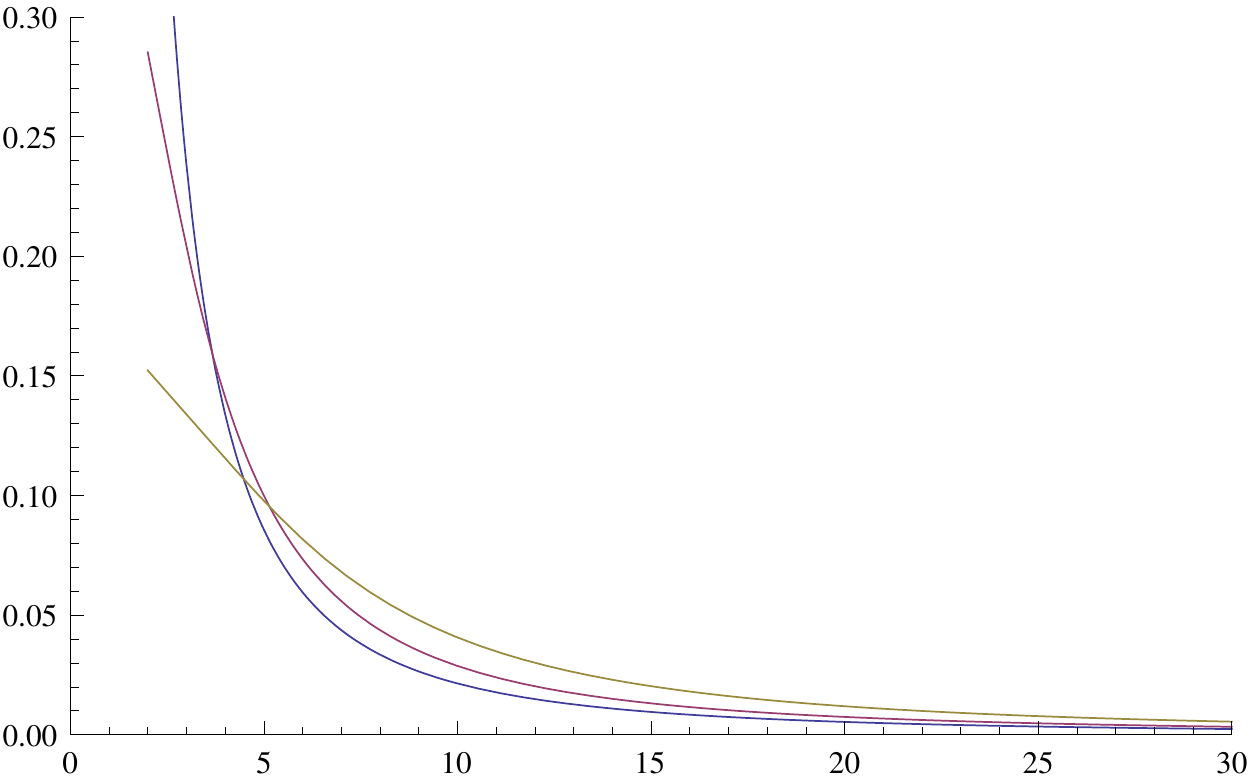}		
		\includegraphics[trim=0cm 0cm 0cm 0cm, clip=true,width=7cm]{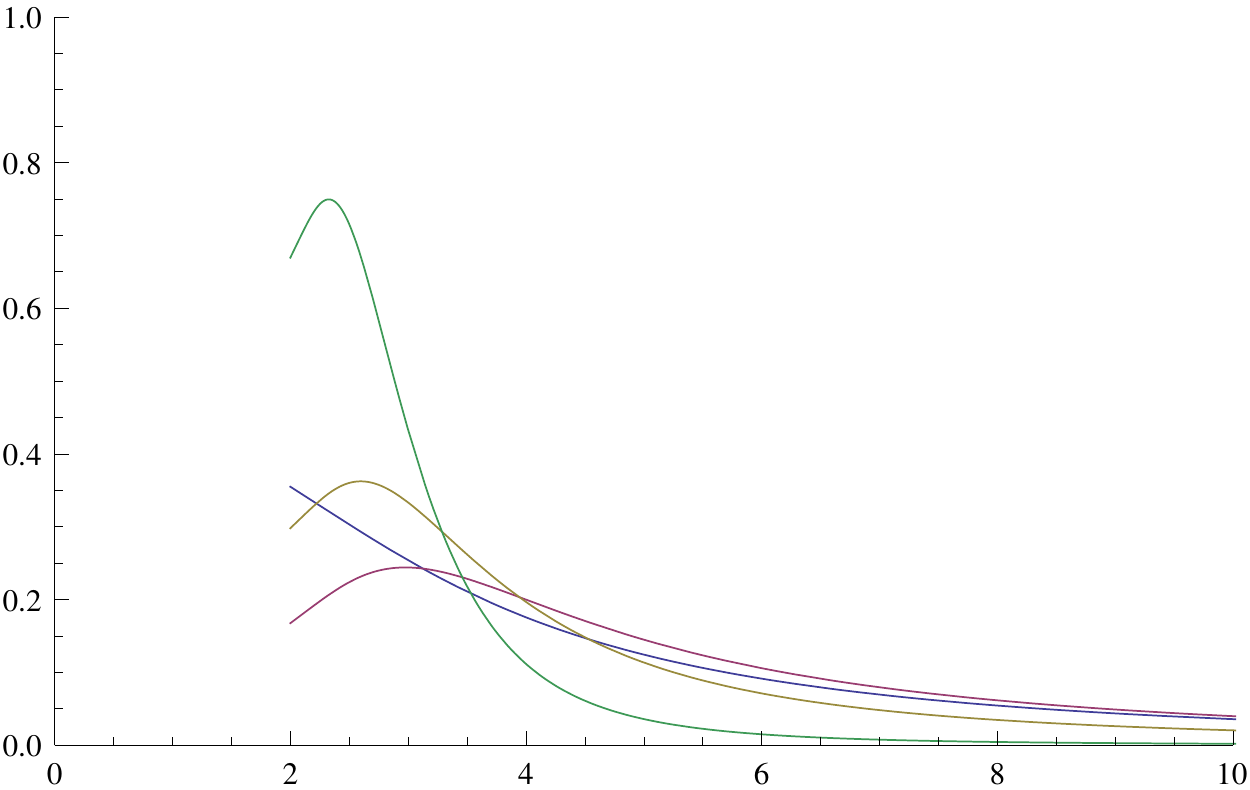}
	\caption{ Examples of $p(\Lambda|d)$ and $p^*(\Lambda|d)$ distributions for the BSM coin, assuming  various data.
	Left pannel: $p(\Lambda|d)\times \Lambda$ (top) and $p^*(\Lambda|d)\times \Lambda$ (bottom) distributions for $(H,T)=(1,1)$, $(5,5)$, $(20,20)$ in respectively blue, purple, yellow. 
		Right pannel: $p(\Lambda|d)\times \Lambda$ (top) and $p^*(\Lambda|d)\times \Lambda$ (bottom) distributions for $(H,T)=(5,5)$, $(5,15)$, $(5,20)$, $(5,30)$ in respectively blue, purple, yellow, green.
	 \label{fig:BSM_coin}}
\end{figure}

As a final illustration, $p(\Lambda|d)$ and $p^*(\Lambda|d)$ are shown on Fig. \ref{fig:BSM_coin} for various outcomes of the BSM
 coin tossing.  As discussed in Sec. \ref{se:EFT_inference}, the shapes remain roughly  identical when data are compatible with the SM. In contrast,  a bump  appears in $p(\Lambda|d)$    when the data favor the BSM hypothesis. The high-$\Lambda$ tail of
  $p^*(\Lambda|d)$ drops increasingly quick    with the increase of BSM evidence.
 
\end{document}